# Experimental demonstration of an ultra-compact on-chip polarization controlling structure

Qingzhong Deng, Lu Liu, Zhiping Zhou*
*State Key Laboratory of Advanced Optical Communication Systems and Networks,
School of Electronics Engineering and Computer Science, Peking University, Beijing ,China 100871*
*\*zjzhou@pku.edu.cn*

**Abstract:** We demonstrated a novel on-chip polarization controlling structure, fabricated by standard 0.18-μm foundry technology. It achieved polarization rotation with a size of 0.726 μm × 5.27 μm and can be easily extended into dynamic polarization controllers.

## 1. Introduction

Silicon on insulator (SOI) is a prevailing platform for its CMOS compatibility and integration compactness. Simultaneously, strong polarization dependence occurs which makes polarization control essential. A great deal of effort has been made but on-chip polarization control is still on the way mainly due to the fabrication problems, either too small features size (60 nm) [1], stringent fabrication accuracy requirement [2, 3], large device size (longer than 100 μm) [4] or incompatibility with CMOS fabrication process [5].

In this paper, we propose and experimentally demonstrate a practical polarization controlling structure with compact size, wide bandwidth and low insertion loss. Moreover, the structure directly links dynamic polarization control with phase manipulation and it can be easily fabricated by the mature 0.18-μm CMOS technology.

## 2. Polarization rotation (Static polarization control)

A partially etched slot waveguide is introduced to realize the polarization rotation [Fig. 1(a)]. $H$=220 nm, $Slot$=200 nm and $Slab$=90 nm are chosen according to the design rules of the fabrication foundry, while $W$=726 nm is optimized for polarization rotation. As shown in Fig. 1(b), it supports three eigenmodes. The mode with $n_{eff}$ =2.222, marked as $TE_0^{slot}$, is quasi-TE polarized since the major transverse fields are $E_x$ and $H_y$. The other two modes, $EM_1^{slot}$ and $EM_2^{slot}$, are hybridized in polarizations.

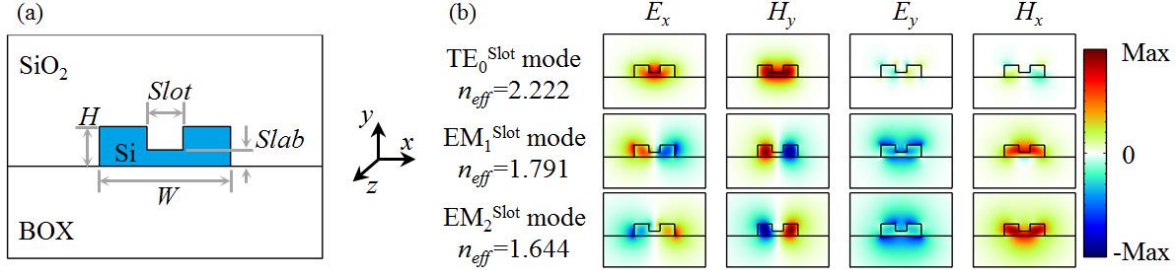

Fig. 1. (a) Schematic and (b) Eigenmodes of partially etched slot waveguide. All simulations are done by 3D full vector finite-element- method (FEM) in this paper. The refractive indices of Si and SiO$_2$ are set to 3.48 and 1.45, respectively.

When a fundamental TM mode in strip waveguide ($TM_0^{strip}$) is injected into such a partially etched slot waveguide, $EM_1^{slot}$ and $EM_2^{slot}$ will be excited since it does not support $TM_0^{slot}$ mode. The effective refractive index ($n_{eff}$) difference between $EM_1^{slot}$ and $EM_2^{slot}$ modes will cause beat and $W$ has been optimized to achieve polarization rotation at half-beat length $L_\pi = \lambda / (2\Delta n_{eff}) = 5.27$ μm, as shown in Fig. 2(a). Then, two strip waveguides are utilized to symmetrically decouple the light into two separate $TE_0^{strip}$ modes with a phase difference of π. When $TE_0^{strip}$ is injected, it will be converted to $TE_0^{slot}$ due to the field similarity and decoupled into two separate $TE_0^{strip}$ modes with a phase difference of 0. This polarization rotation scheme has been verified with the measured optical images in Fig. 2(b). Inputting $TE_0^{strip}$ and $TM_0^{strip}$ into this structure simultaneously, interference occurs in the two output ports and a wavelength shift of FSR/2 is observed [Fig. 2(c)]. It can be explained as: in a reference waveguide (WG$_{ref}$), $TE_0^{strip}$ and $TM_0^{strip}$ modes will not interfere with each other since they are orthogonal. However, if the injected $TM_0^{strip}$ mode gradually transforms into $TE_0^{strip}$ mode and it has a phase delay with the original $TE_0^{strip}$ mode, interference will take place and result in two complementary output transmission spectra. From this perspective, the observed interference and wavelength shift are also strong evidences of polarization rotation.

Separating the incident optical power in $TE_0^{strip}$ and $TM_0^{strip}$ modes is always wanted. A polarization rotator and splitter (PRS) is proposed by attaching a π/2 phase-shifter and a directional coupler (DC) [Fig. 3(a)]. The measured results in Fig. 3(b) and 3(c) indicates that the insertion loss (IL) is -0.90 dB (-2.28 dB) and the extinction ratio (ER) is 14.56 dB (11.77 dB) at 1550 nm wavelength for $TE_0^{strip}$ ($TM_0^{strip}$) incidence. The insets in Fig. 3(b) and 3(c) indicates that the output light is dominated by TE polarization but have not been fully coupled into one waveguide. Therefore, the performance will be better if the DC is optimized properly. Moreover, simulations proved that using strip-slot mode converter [6] can further elevate the performance: the IL is -0.7 dB (-0.9 dB) and the ER is 22 dB (34 dB) for $TE_0^{strip}$ ($TM_0^{strip}$) incidence.

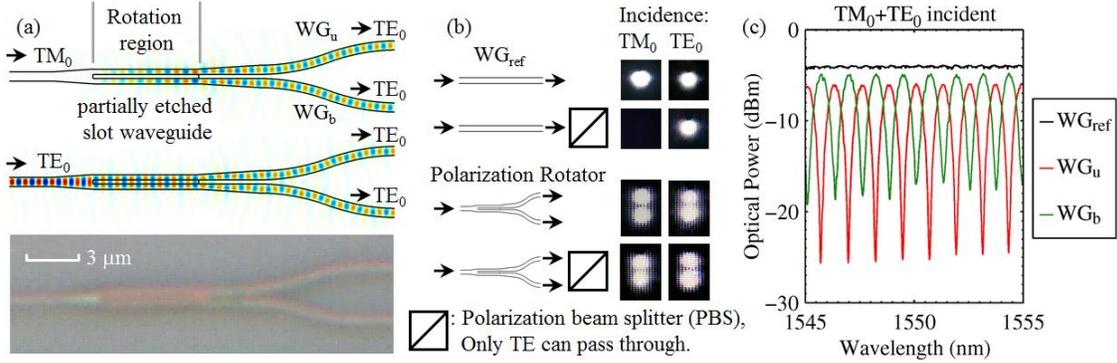

Fig. 2. (a) Optical field evolution ($H_y$) and Micro micrographs of the proposed polarization rotator (PR); (b) optical images captured from the output waveguides; (c) The measured transmission spectra for hybrid $TE_0$ and $TM_0$ incidence.

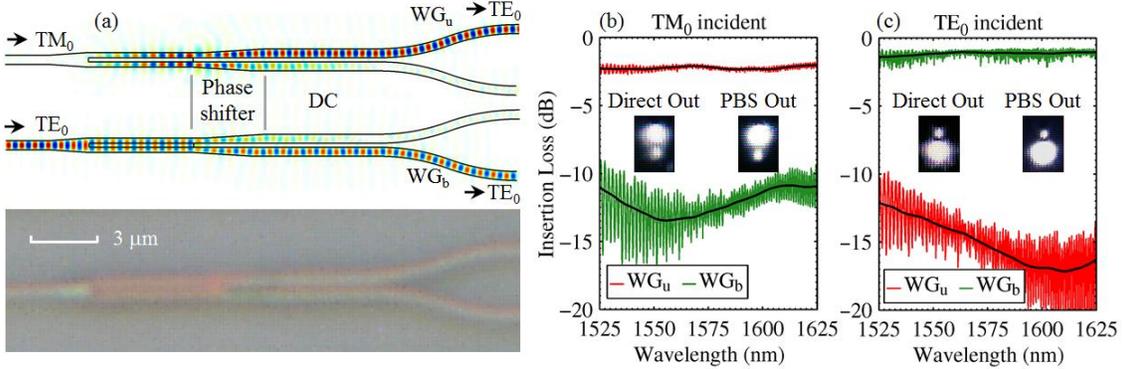

Fig. 3. (a) Optical field evolution ($H_y$) and Micro micrographs of the proposed polarization rotator and splitter (PRS); the measured transmission spectra and optical images (Insets): (b) $TM_0$ incident and (c) $TE_0$ incident.

### 3. Dynamic polarization control

As discussed above, the proposed polarization rotation structure associates the phase difference with polarization state: $TE_0$ corresponds to two light beams with phase difference of 0 and $TM_0$ corresponds to π. Here we show an example of dynamic polarization controllers by integrating the structure with phase tuning elements (Fig. 4).

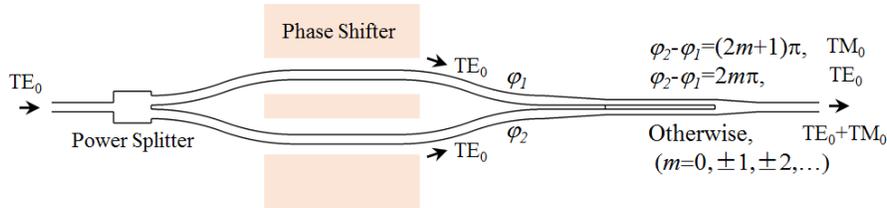

Fig. 4. Schematics of dynamic polarization controller.

### 4. Conclusion

We have proposed and designed an untra-compact polarization rotation structure using partially etched slot waveguide which is just 0.726 μm × 5.27 μm in size and can be easily fabricated. The function of rotation is experimentally verified with a PR and PRS which are fabricated by standard 0.18-μm foundry technology. Moreover, the structure associates polarization states with phase difference which enables on-chip dynamic polarization control by mature phase manipulation in silicon photonics.

This work was supported by the Natural Science Foundation of China (Grant No. 61120106012).

**Reference:**
[1] A. V. Velasco, M. L. Calvo, and P. Cheben, et. al., "Ultracompact polarization converter with a dual subwavelength trench built in a silicon-on-insulator waveguide," Opt. Lett. **37**, 365 (2012).
[2] L. Liu, Y. Ding, K. Yvind, and J. Hvam, "Efficient and compact TE-TM polarization converter built on silicon-on-insulator platform with a simple fabrication process," Opt. Lett. **36**, 1059-1061 (2011).
[3] L. Gao, Y. Huo, and K. Zang, et. al., "On-chip plasmonic waveguide optical waveplate," Sci. Rep.-UK **5**, 15794 (2015).
[4] W. D. Sacher, T. Barwicz, B. J. F. Taylor, and J. K. S. Poon, "Polarization rotator-splitters in standard active silicon photonics platforms," Opt. Express **22**, 3777-3786 (2014).
[5] H. Deng, D. Yevick, C. Brooks, et. al., "Design Rules for Slanted-Angle Polarization Rotators," J. Lightwave Technol. **23**, 432 (2005).
[6] Q. Deng, Q. Yan, and L. Liu, et. al., "Robust polarization-insensitive strip-slot waveguide mode converter based on symmetric multimode interference," Opt. Express **24**, 7347-7355 (2016).